\documentclass[aps,pre,epsfig,preprint]{revtex4}
\usepackage{epsfig}
\begin{document}
\title{Boltzmann temperature in out-of-equilibrium lattice gas}
\author{M\'ario Jos\'e de Oliveira and Alberto Petri}
\affiliation{$^{(1)}$Instituto de F\'{\i}sica, Universidade de S\~ao Paulo, 
Caixa Postal 66318, 05315-970 S\~ao Paulo, S\~ao Paulo, Brazil\\  
$^{(2)}$CNR - Istituto dei Sistemi Complessi, via del Fosso del Cavaliere 100,
00133 Roma, Italy}
\begin{abstract}
We investigate  the quench of Ising and Potts models via Monte Carlo dynamics, 
and find that the distribution of the site-site interaction energy
has the same form as in the equilibrium case. 
This form directly derives from the Boltzmann statistics and 
allows to measure the 
instantaneous temperature during the systems relaxation. 
We find that, after an undercritical quench, the system equilibrates in a finite
time at the heatbath temperature, while the energy still decreases due to the 
coarsening process. 
\end{abstract}
\maketitle
\section{Introduction}

A key quantity in Monte Carlo simulation of equilibrium systems
is the heathbath temperature $T$. 
Monte Carlo dynamics follows 
a Markovian process
whose stationary state is proportional to the Boltzmann
probability density at the chosen temperature, i.e.:
\[
P(E) \propto \exp[-\beta E],
\]
where $E$ is the system energy and $\beta$ is the inverse heathbath temperature
in Boltzmann's constant units. Temperature  is usually a parameter with
an imposed value, the fluctuating quantity being the energy, and this corresponds 
to simulate a canonical ensemble. It has been recently shown \cite{shida03}
that also a microcanonical formulation can be constructed, in which 
the system energy is constrained whereas the temperature is
computed as a dependent quantity. In this formulation the temperature is 
found to assume the correct value at equilibrium, 
thus establishing the equivalence of 
simulating the microcanonical and the canonical ensemble. 
 
Our work aims to investigate the possibility of associating an istantaneous
temperature to a system even when this is out of the quilibrium.
Therefore we have employed the method developed in Ref.~\cite{shida03}
in a non equilibrium case, by computing the temperature of the Ising model along a quench from high 
temperature. We have then extended the method to 
the Potts model with $q$ states.
With similar aims different approaches for off-lattice systems 
(molecular dynamics) have been recently introduced 
\cite{rugh97,morriss99,han04} 
that however are not suitable for lattice models.

In the method adopted here computation bases  
on the statistical distribution of the site-site interaction 
energies. 
The most interesting finding is that the shape of the 
distribution
computed  along the quench is very similar to the one
which characterizes the system at equilibrium,  but corresponds to a 
different value of the temperature.
Thence  
it is possible to associate an instantaneous temperature 
to the non equilibrium states of the system.

In Section II we recall how temperature can be determined in the Ising system
at equilibrium; in Sec. III the method is applied to the same system in 
non equilibrium and in Sec. IV it is extended to the Potts model 
with $q$ states. Main results are summarized in Sec. V.

\section{The Ising model}

In Ref. \cite{shida03} the equilibrium temperature 
is computed 
on the  base of the statistical properties of the site-site interaction 
energies for a generic 
lattice gas model. In the case of the Ising model, with Hamiltonian
\begin{equation}
\label{ising}
H = - \sum_{\langle i<j\rangle} \sigma_i \sigma_j,
\end{equation}
where the sum $\langle \dots \rangle$ is over nearest neighbors sites
and $\sigma_k =\pm 1$,
the temperature is obtained from the distribution of the quantity:
\[
\gamma = - \sum_{\left[v\right]}    \sigma_0 \sigma_v ,
\]
that represents the interaction energy of a generic site $k=0$ 
with its neighbors $\left[v\right]$. 
The Hamiltonian, Eq.~(1), can be separated into $H=\gamma+{\cal H}_r$, where
\[
{\cal H}_r=- \sum_{\langle l<m \rangle} \sigma_l \sigma_m,
\]
with the the sum extended to all pairs not including the site $0$.
Say $P({\{\boldmath{\sigma}}\})$ the (Boltzmann) probability  
for the system of $N+1$ sites to stay in the configuration 
$\{\boldmath{\sigma}\}=\{\sigma_0,\sigma_1,\dots,\sigma_N\}$, this
is given by
\[
P({\{\boldmath{\sigma}}\})=\frac{1}{\cal{Z}} 
e^{-\beta \gamma} e^{- \beta {\cal H}_r},
\]
where $\beta$ is the inverse temperature (in Boltzmann's constant units) 
and  $\cal{Z}$  the partition function.
The probability $P(\epsilon)$ of a configuration with $\gamma=\epsilon$,
$(\epsilon=-4,-2,0,2,4)$ can be obtained as
\begin{eqnarray*}
P(\gamma=\epsilon)=\sum_{\{\boldmath{\sigma}\}} \delta(\epsilon,\gamma)
P(\{\boldmath{\sigma}\}) =
\sum_{\{\boldmath{\sigma}\}} \delta(\epsilon,\gamma) 
\exp[-\beta\gamma]  \exp[-\beta {\cal H}_r] =\\
\exp[-\beta\epsilon] \sum_{\sigma_0}\sum_{\{\boldmath{\sigma_r}\}} 
\delta(\epsilon,\gamma)\exp[-\beta {\cal H}_r],
\end{eqnarray*}
where $\delta(\alpha,\beta)=1 $ if $\alpha=\beta$ and $=0$ otherwise, and
${\{\boldmath{\sigma_r}\}}$  is
the  set of all possible configurations of the system with $\sigma_0$ fixed.
All and only those configurations 
yielding $\gamma=\epsilon$ contribute to the sums in the last term, 
so they add to a number which only may depend
on the temperature and $\epsilon$, thus:
\[
P(\gamma=\epsilon) = e^{-\beta\epsilon}  a(\beta,\epsilon).
\]
In order to avoid the explicit computation of $a(\beta,\epsilon)$, which
is as difficult as that of the partition function, let us consider:
\begin{eqnarray*}
P(\gamma=-\epsilon)=\sum_{\{\boldmath{\sigma}\}} \delta(-\epsilon,\gamma)
P(\{\boldmath{\sigma}\}) =
\sum_{\{\boldmath{\sigma}\}} \delta(-\epsilon,\gamma)
\exp[-\beta\gamma]  \exp[-\beta {\cal H}_r] =\\
\exp[\beta\epsilon] \sum_{\sigma_0}\sum_{\{\boldmath{\sigma_r}\}}
\delta(-\epsilon,\gamma)\exp[-\beta {\cal H}_r].
\end{eqnarray*}
By letting $\sigma_0 \rightarrow \sigma_0'= -\sigma_0$, thus
$\gamma \rightarrow -\gamma$, and the last expression yields
\[
P(\gamma=-\epsilon)=\exp[\beta\epsilon] \sum_{\sigma_0'}\sum_{\{\boldmath{\sigma_r}\}}
\delta(-\epsilon,-\gamma)\exp[-\beta {\cal H}_r],
\]
that is, since $\delta(-\epsilon,-\gamma)=\delta(\epsilon,\gamma)$,
\[
P(\gamma=-\epsilon)=e^{\beta\epsilon} a(\beta,\epsilon).
\]
Thence the ratio
\[
\frac{P(\gamma=\epsilon)}{P(\gamma=-\epsilon)}=e^{-2\beta\epsilon},
\]
does not depend on $a(\epsilon,\beta)$.

This equation derives from the canonical distribution and 
can be used to estimate the lattice temperature 
$T_B$ in numerical simulations of the Ising model since, by the law of the
large numbers, one can approximate
\begin{equation}
\label{erre}
\frac{P(\gamma=\epsilon)}{P(\gamma=-\epsilon)}\simeq
\frac{n(\gamma=\epsilon)}{n(\gamma=-\epsilon)}=R,
\end{equation}
being $n(\gamma=\epsilon)$ the number of lattice sites with 
$\gamma=\epsilon$, and thence
\begin{equation}
\label{temperature}
r=T_B \cdot \epsilon
\end{equation}
with
\[
r=-2\left(\ln R \right)^{-1}.
\]
If necessary, better estimates  can be obtained by computing averages
of  $n(\epsilon,\beta)$ over time, since the system is at equilibrium.
In Ref. \cite{shida03} it has been checked that 
Eq.~(\ref{temperature}) is well satisfied when simulating the 
canonical ensemble, and has been used to compute temperature in 
a microcanonical Monte Carlo simulation with the Kawasaki 
\cite{reviewmc} dynamics.
\hfill\break

\section{Non equilibrium Ising model}

Monte Carlo simulations are 
extensively employed even for studying non equilibrium properties of systems
\cite{reviewmc}.
Among the others, the properties of systems prepared far from
equilibrium and then allowed to relax towards their equilibrium state.
If for instance the system is initially at high 
temperature and is put then in contact with a finite temperature 
heat bath, it starts to cool down and eventually approaches the heath bath 
temperature.
Despite the 
intrinsic artificial nature of Monte Carlo dynamics, 
physical relevance is
generally attributed to results obtained via single site dynamics
\cite{reviewmc}. 
It is therefore tempting to use Eq.~(2) for probing the system
temperature during the cooling. One wishes, in particular, to
answer the
following two questions:
\begin{itemize}
\item[1)] is the system  at equilibrium on very short time scales? i.e.
does  Eq.~(\ref{temperature}), which was derived by
assuming the Boltzmann (equilibrium) statistics, hold during the system cooling?
\item[2)] if yes, how does temperature decrease in time during the cooling?
\end{itemize}

In order to answers the above questions we have performed
Monte Carlo simulations in which the $2d$ Ising model was quenched from
infinite to finite temperature.  
We have found that the answer to question $1)$ is affirmative.

In Fig.~1, $r(\epsilon)$ 
is shown for a quench 
from $T=\infty$ 
to $T=3.0$ at some different times. 
In this case the 
quench temperature is above  $T_c$, and  a few Monte Carlo steps
are sufficient for driving the system to thermal equilibrium. However, 
before this happens the system  already displays a linear relation between 
$\epsilon$ and $r$, from which an istantaneous temperature can be derived.
as seen also in the inset where 
$r/\epsilon$ is shown.  A linear regression of $r$ vs $\epsilon$ at 
equilibrium yields $T=2.996$, with  correlation coefficient $=1$.

A subcritical quench is shown in Fig~2. Here $T=2.0$. In this 
case the system takes a long time to reach thermal equilibrium, but 
Eq.~(\ref{temperature}) is quite well satisfied. 
It can be seen from the inset that deviations from the linear
dependence exist. They are however small
for any  $\epsilon$, and their average is zero.
Quasi-equilibrium is satisfied at any instant and in this case a 
linear regression on the last curve gives $T_B=2.058$ with a correlation
coefficient $=0.9998$.

The existence of such an equilibrium-like distribution for the system allows for
extracting a Boltzmann temperature $T_B$ during the system coooling. 
Figure~3 shows the behavior of
the energy as function of this temperature for a quench at $T=2.0$, i.e. 
below the critical
temperature  $T_c \simeq  2.26$.
It is seen that 
a first regime exists in which the two quantities are proportional.
Then, a second regime is entered where a
remarkable property can be noticed: 
the Boltzmann temperature attained that of the heatbath but, 
on the contrary, the energy is still relaxing.
In fact this kind of systems is characterized by a coarsenig process and
takes an infinite time to go to equilibrium, the energy decaying as
$1/\sqrt(t)$, but it is seen that temperature is already at equilibrium 
during the coarsening.
This behaviour is better pointed out in the inset, 
where it is seen that energy decreases algebraically while $T_B$ is
constant.

When the system is quenched at a temperature  above $T_c$, 
energy relaxes exponentially
to equilibrium and the second regime does not appear.
This is shown in Fig.~4 and in the related inset.
For comparison, the equilibrium curve is also reported in the figures, 
showing that it still lies below the non-equilibrium one.

\section{The Potts model}

In order to test the above conclusions on a wider class of systems, we have 
derived Eq.~(\ref{temperature}) for the case of the $q$ states
Potts model, with Hamiltonian: 
\begin{equation}
\label{potts}
H = \sum_{\langle i<j \rangle} (1-\delta_{\eta_i \eta_j})
\end{equation}
where again the sum is over the nearest neighbors, $\delta_{\alpha \beta}$
is the Kronecker's function, and $\eta_k$ is the state of site $k$.
Each site can assume one out of $q$ different states, so that one can 
identify the state with an integer: $\eta_k =1,2,\dots,q$ and 
the system energy Eq.~(\ref{potts}) can be written as 
\[
E(\eta) = \gamma + {\cal H}_r,
\]
where
\begin{equation}
\label{inten}
\gamma = \sum_{v} (1-\delta_{\eta_0 \eta_v})
\end{equation}
is the interaction energy of the site in $0$ with its neighbors, and
\[
{\cal H}_r= \sum_{\langle i<j \rangle} \xi(\eta_i,\eta_j)
\]
is the energy interaction of the rest of the system.

Let us now define $\gamma_{\rho}$ the energy Eq.~(\ref{inten}) 
when $\eta_0=\rho$ and similarly $\gamma_{\omega}$  the same quantity 
when $\eta_0=\omega$.
In analogy with the case of the Ising model:
\begin{equation}
\frac{P(\rho,\eta_1,\dots,\eta_N)}{P(\omega,\eta_1,\dots,\eta_N)}
=
\exp[-\beta (\gamma_\rho-\gamma_\omega)]=\exp[-\beta \gamma_{\rho \omega}].
\end{equation}
Thus the probability for $\gamma_{\rho \omega}$ to assume the value $\epsilon$
is given by
\begin{eqnarray*}
\sum_{\eta_1,\dots,\eta_N} \delta(\epsilon,
\gamma_{\rho\omega})P(\rho,\eta_1,\dots,\eta_N) =\\
  \sum_{\eta_1,\dots,\eta_N} \delta(\epsilon,\gamma_{\rho\omega})
\exp[-\beta\gamma_{\rho\omega}]  P(\omega,\eta_1,\dots,\eta_N) =\\
\exp[-\beta\epsilon] \sum_{\eta_1,\dots,\eta_N}
\delta(\epsilon,\gamma_{\rho\omega}) P(\omega,\eta_1,\dots,\eta_N).
\end{eqnarray*}
By multiplying the left hand side by $1=\sum_{\eta_0} \delta(\eta_0,\rho)$
and he right hand side by $1=\sum_{\eta_0} \delta(\eta_0,\omega)$
one gets
\[
\sum_{\{\boldmath{\eta}\}} \delta(\eta_0,\rho) 
\delta(\epsilon,\gamma_{\rho\omega}) P(\{\eta\}) =
e^{-\beta \epsilon} \sum_{\{\boldmath{\eta}\}} \delta(\eta_0,\omega) 
\delta(\epsilon,\gamma_{\rho\omega})P(\{\eta\}),
\]
where $\{\eta\}$ is the system configuration, and thus finally
\begin{equation}
\exp[-\beta \epsilon] =
\frac{\langle \delta(\eta_0,\rho) \delta(\epsilon,\gamma_{\rho\omega}) 
\rangle}
{\langle \delta(\eta_0,\omega) \delta(\epsilon,\gamma_{\rho\omega}) \rangle}.
\end{equation}
Thence even for the Potts model the temperature can be derived from the
site-site energy statistics. 
The same argument as above holds for  
any pair of different states $\eta_0^{{(\alpha)}},\eta_0^{(\beta)}$, 
so that one can also write  
\begin{equation}
\label{pottstemperature}
\exp[-\beta \epsilon] = \frac{1}{q(q-1)}
\sum_{\eta_0^{(\alpha)},\eta_0^{(\beta)}}
\frac{\langle \delta(\eta_0,\eta_0^{(\alpha)}) 
\delta(\epsilon,\gamma_{\eta_0^{(\alpha)} \eta_0^{(\beta)}}) 
\rangle}
{\langle 
\delta(\eta_0,\eta_0^{(\beta)}) \delta(\epsilon,
\gamma_{\eta_0^{(\alpha)} \eta_0^{(\beta)}})
\rangle}.
\end{equation}
This can be helpful for improving statistics in out of equilibrium simulations 
where the thermal average cannot be obtained  by averaging over time like 
in equilibrium cases. One thus replaces thermal averages with system  averages
in  Eq.~(\ref{pottstemperature}).

It is found that the Potts model obeys Eq.~(ref{pottstemperature}) aven in the non equilibrium regime, allowing to extract an instantaneous temperature 
$T_B$ like done for the Ising model. 
Figure 5 shows the Boltzmann
temperature $T_B$ vs time for a quench of a seven state Potts model at $T=0.5$, i.e. 
below the critical
temperature  $T_c \simeq  0.77$.
It is seen that like for the Ising model
a first regime exixts in which the two quantities are proportional.
Then  even in this case the second regime is entered in which
$T_B$ is that of the heatbath but
the energy is still relaxing due to the coarsening process.
The supercritical case reported in Fig.~6 shows that
$T_B=T$ after a certain time, as for the Ising model.

\section{Low temperature behaviour} 
We have performed many quenches at different final temperature $T$ for 
the the Potts model with $q=7$ and $q=2$. The latter is equivalent to 
the Ising model by shifting energies by adding $2$ and compressing them by a 
factor $1/2$ ($T-c \simeq 1.13$).  
A common feature is that  as $T$ is decreased below a given value, the 
system  seems unable to reach the equilibrium temperature.
 
Figure~7 shows $T_B$ vs time for the Ising model with linear lattice size 
$L=1000$.
The Boltzmann temperature attains the value of $T$ if
this is above $T_l > 0.3$ but if $T$ is equal or below this value it doesn't. 
We expect that this is a finite size effect and that the system eventually
relaxes to the ground state with $T_B=0$.
 The case of the Potts model shown in Fig.~8. In this case $T_l \simeq 0.3$.
It can  be observed that $T_l \approx 0.42 T_c$ for Potts whereas 
$T_l \approx 0.15 T_c$ for Ising. 
Moreover the limit energy is clearly larger for Potts tha for Ising.
Figure~9  displays  the behaviour of $T_B$ after a quench at $T=0$.
An increase of $T_l$ with increasing size for both Ising and Potts model is
seen. Although extrapolation to $L=\infty$ appears to give a still  finite 
temperature, it must be point out that the evaluation of $T_B$ via 
Eq.~(\ref{erre}) becomes difficult for quenches at very low temperature,
since almost all sites  are in the same state and statistical sampling for 
$\epsilon \ne 0$ becomes very poor.
Reliable estimates of $T_B$ can be obtained only for finite values of the 
quench temperature above a bound which depends on system size and kind.

\section{Summary}

In this work a method for associating an instantaneous temperature to non 
equilibrium lattice models has been presented  which is based on the 
the probability distribution of the site-site interaction energies and 
which is derived from the equilibrium Boltzmann statistics. It is found 
that the same form of distribution is obeyd by the system  while cooling 
after a quench, allowing to associate a Boltzmann-like temperature
to its non equilibrium states. While cooling, the Boltzmann temperature 
decreases until the value of the heath bath is attained, 
both for  supercritical and undercritical quenches. 
The latter case is particularly worth of notice, since there the system 
undergoes a coarsening process and equilibrium is never attained from the 
energetical point of view
The method proofs valid for generic Potts model with $q$ states, provided 
the heath bath temperature is not to low and the statistical sampling 
is effective for the considered system size.  

\section{Acknowledgments}
This work has been done with the support of CCINT-USP.


\begin{figure}
\centering\epsfig{file=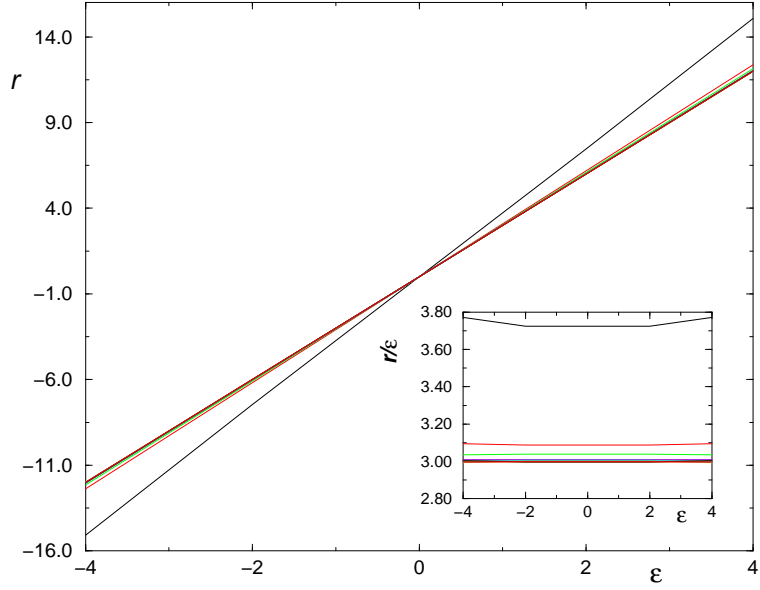, width=8cm, height=10cm, angle=-90}
\caption{\label{fig1}
The distribution of $r(\epsilon)$, Eq.~(\ref{temperature}), for the Ising 
model after a supercritical quench at different times. 
In the inset the quantity $r(\epsilon)/\epsilon)$ is shown (see text).
The time interval between curves is of $5$ mcs.}
\end{figure}

\begin{figure}
\centering\epsfig{file=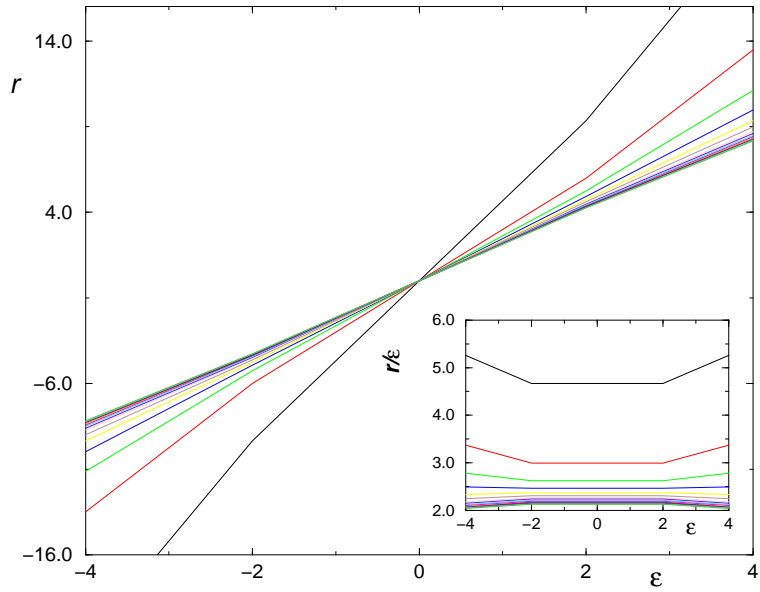, width=8cm, height=10cm, angle=-90}
\caption{\label{fig2}
The distribution of $r(\epsilon)$, Eq.~(\ref{temperature}), for the Ising 
model out of equilibrium. In the inset the quantity $r/\epsilon$ is shown.
The time interval between curves is of $5$ mcs.}
\end{figure}

\begin{figure}
\centering\epsfig{file=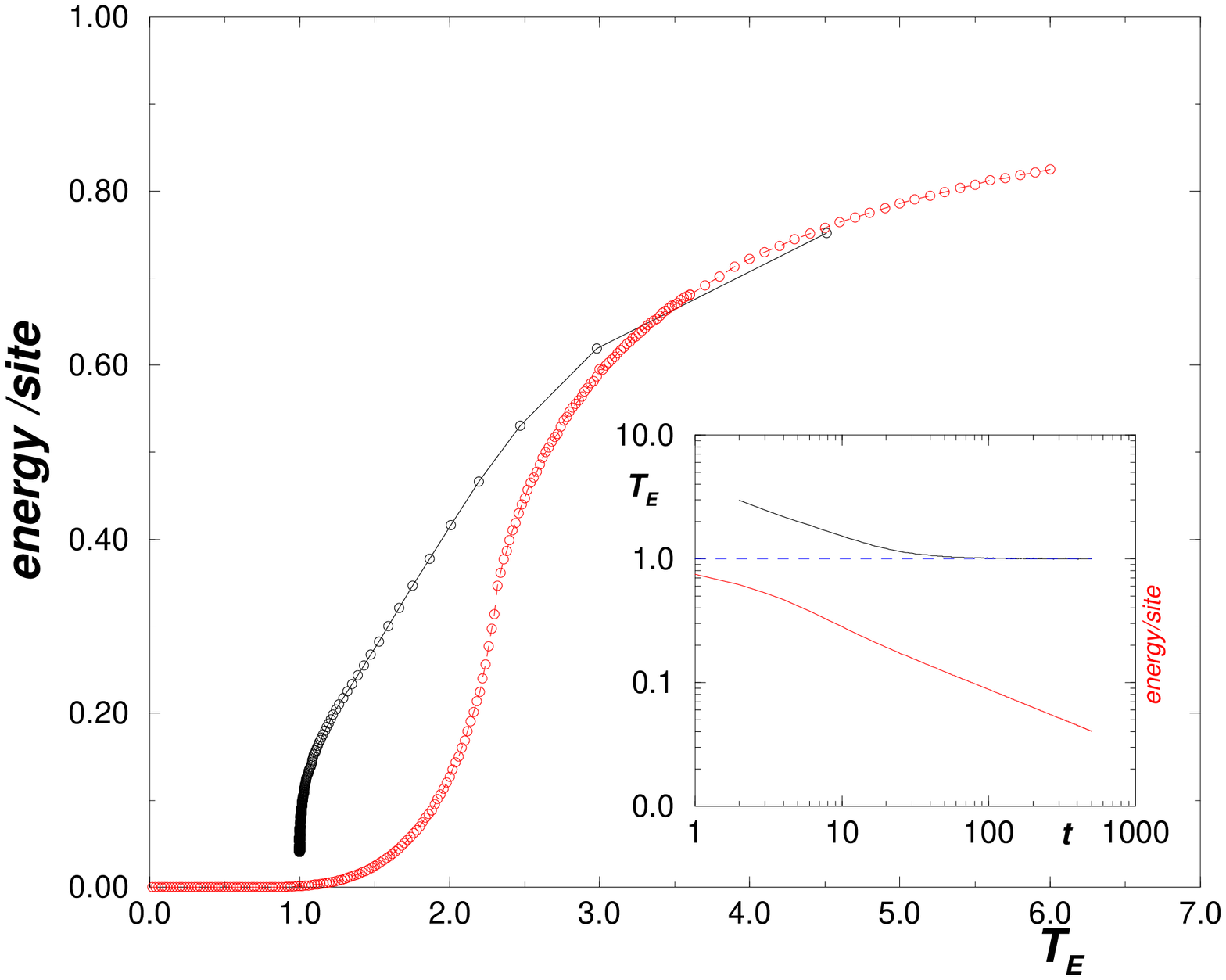, width=8cm, height=10cm, angle=-90}
\caption{\label{fig3}
Energy vs Temperature  during a quench of the Ising model at
$T < T_c$ (black curve). The red dashed curve is the equilibrium curve.
Inset: energy and $T_B$ vs time.}
\end{figure}

\begin{figure}
\centering\epsfig{file=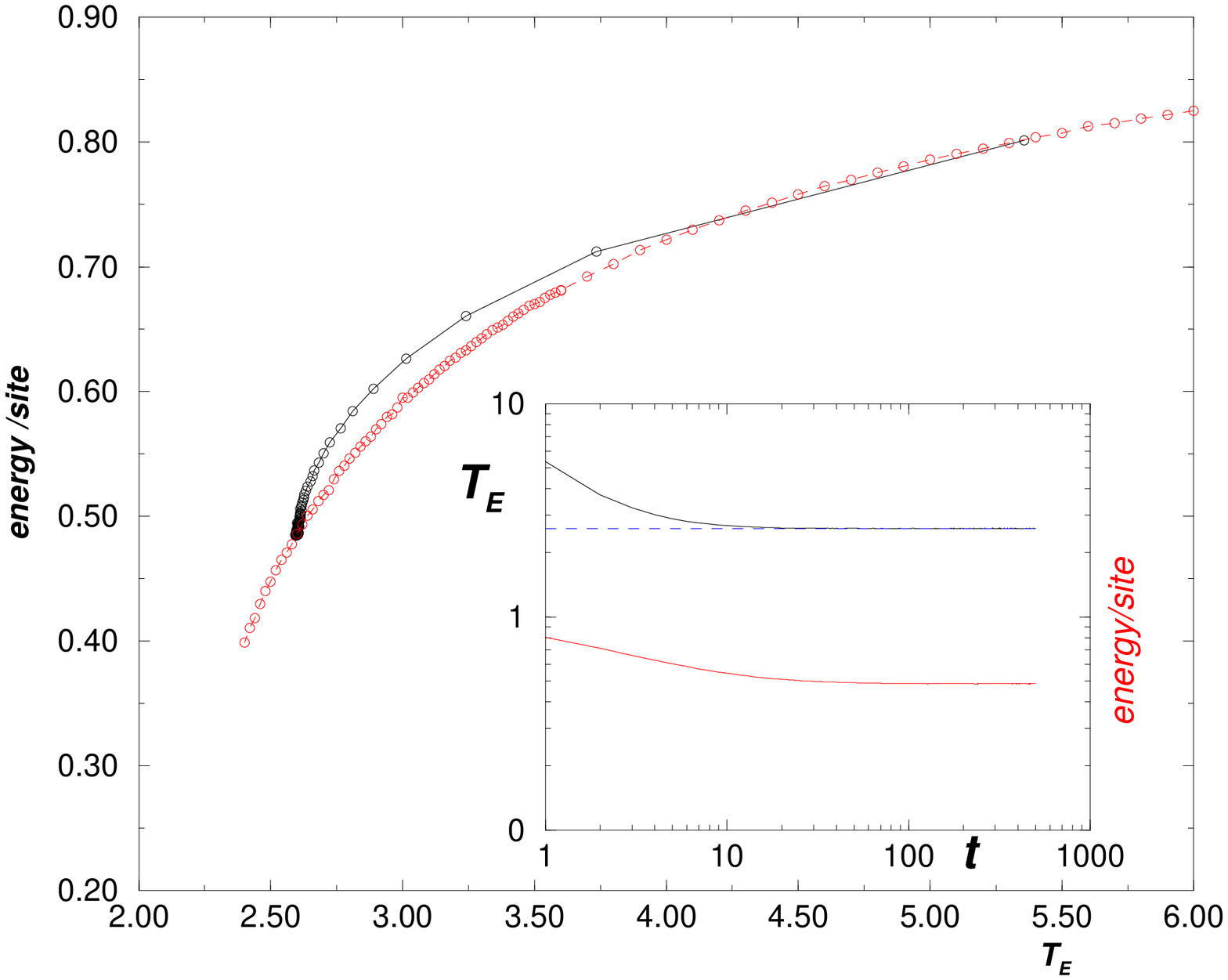, width=8cm, height=10cm, angle=-90}
\caption{\label{fig4}
Energy vs Temperature  during a quench of the Ising model at
$T > T_c$ (black curve). The red dashed curve is the equilibrium curve.
Inset: energy and $T_B$ vs time.} 
\end{figure}

\begin{figure}
\centering\epsfig{file=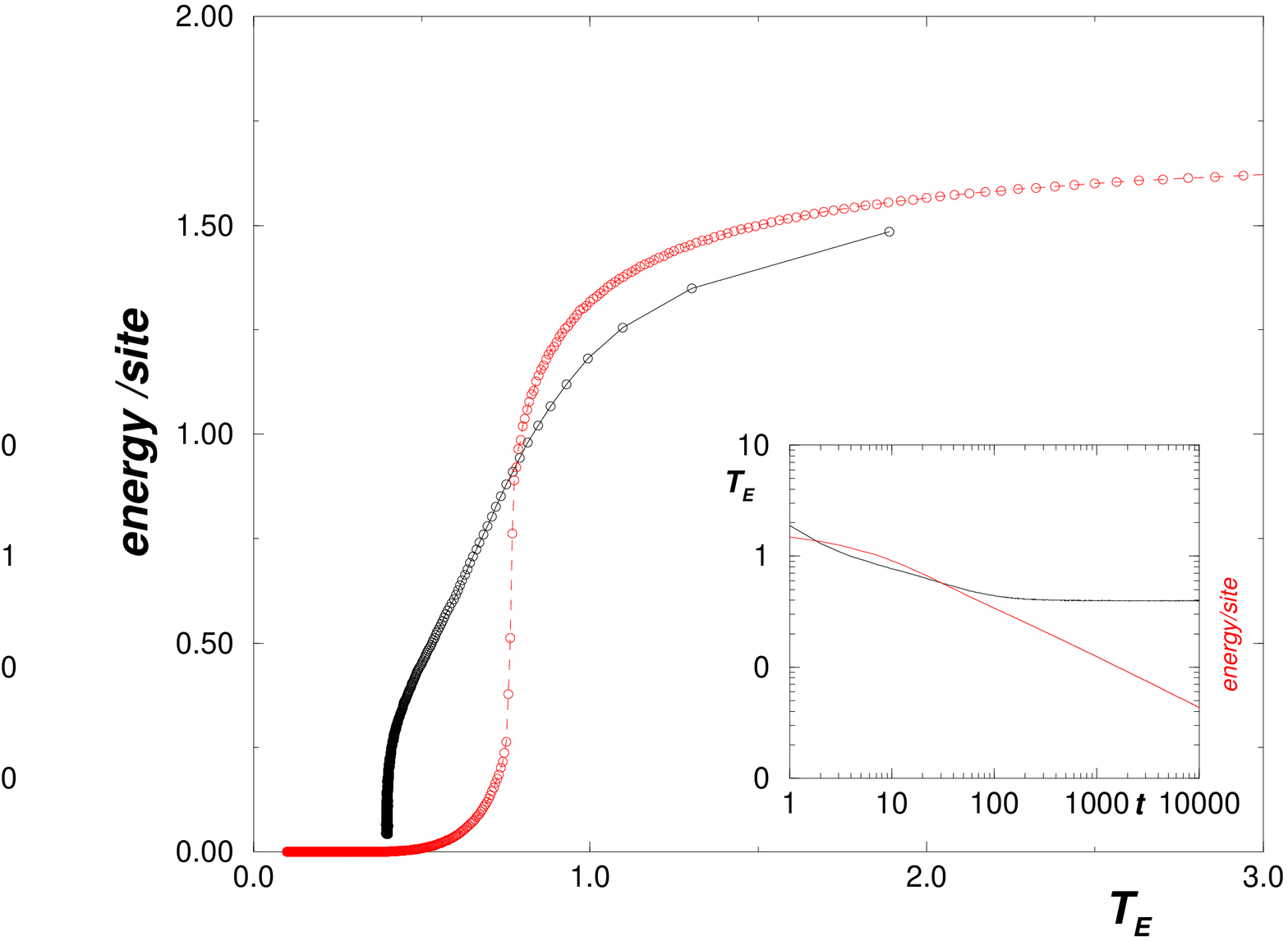, width=8cm, height=10cm, angle=-90}
\caption{\label{fig5}
Energy vs Temperature  during a quench of the $7$ state Potts model at
$T < T_c$ (black curve). The red curve is the equilibrium curve. 
The inset shows $T_B$ 
and energy behavior in time.}
\end{figure}

\begin{figure}
\centering\epsfig{file=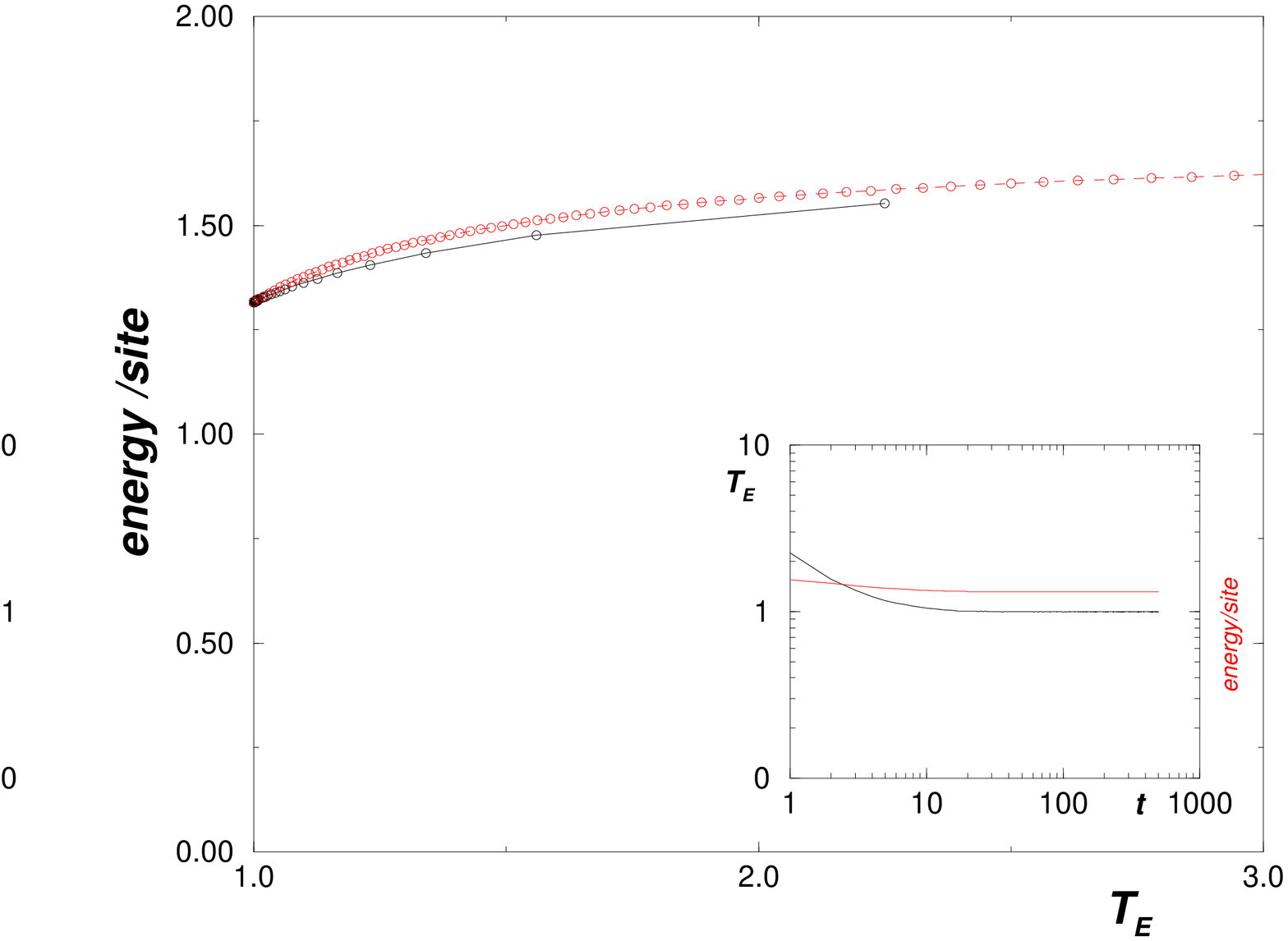, width=8cm, height=10cm, angle=-90}
\caption{\label{fig6}
Energy vs Temperature  during a quench of the $7$ state Potts model at
$T > T_c$ (black curve). Red curve is the equilibrium curve. 
 In the inset, $T_B$ and energy vs time.}
\end{figure}

\begin{figure}
\centering\epsfig{file=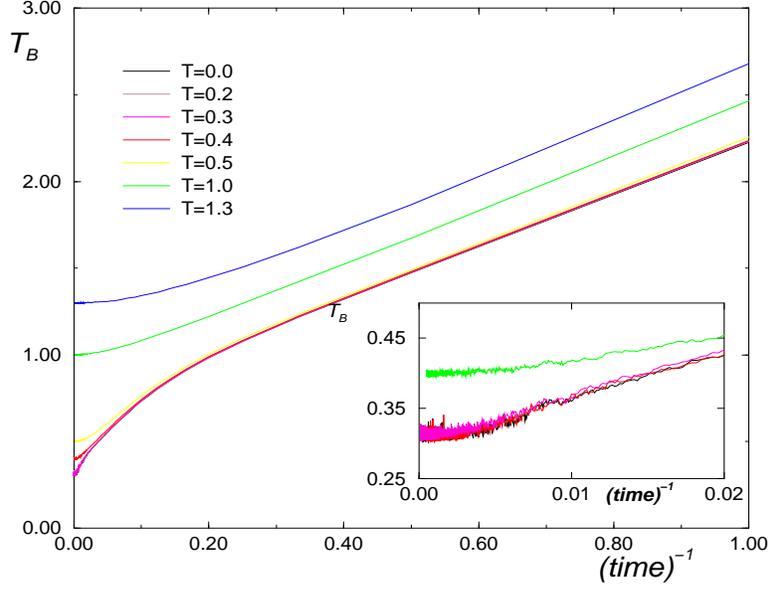, width=8cm, height=10cm, angle=-90}
\caption{\label{fig7}
$T_B$  vs time for the Ising model for different quench temperatures
$T$; the behaviour at very low temperatures is magnified in the inset.
Data are averages over 10 realisations.}
\end{figure}

\begin{figure}
\centering\epsfig{file=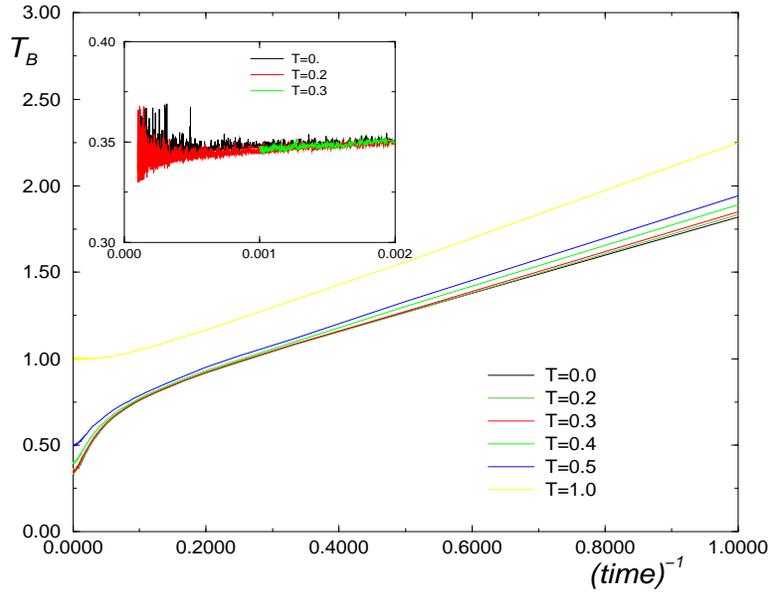, width=8cm, height=10cm, angle=-90}
\caption{\label{fig8}
$T_B$ vs time in  the Potts model for different quench temperatures
$T$; in the inset the behaviour at very low temperatures.Data are averages 
over 10 realisations.}
\end{figure}

\begin{figure}
\centering\epsfig{file=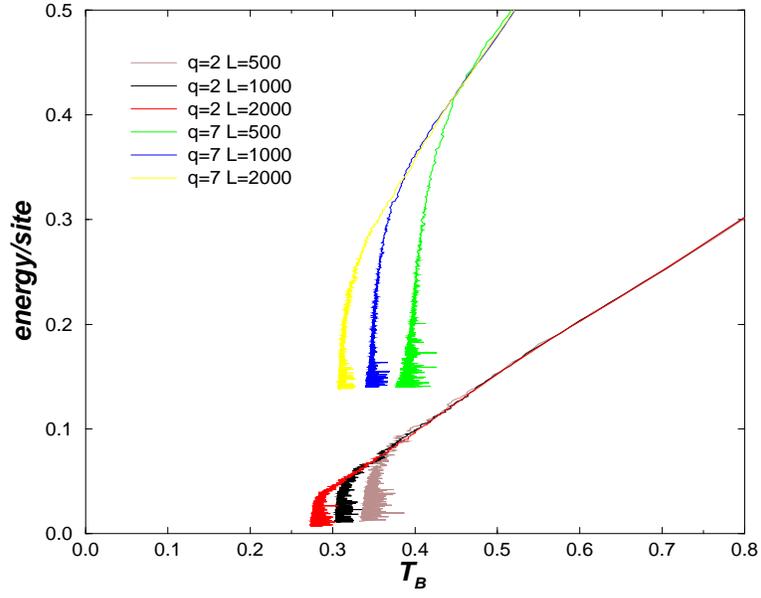, width=8cm, height=10cm, angle=-90}
\caption{\label{fig9}
Energy vs Boltzmann temperature for a quench of the Ising and Potts models 
at $T=0$ in lattices of different size.}
\end{figure}
\end{document}